\documentclass{amsart}

\usepackage{amssymb, amscd,stmaryrd,setspace,hyperref,color}

\input xy
\xyoption{all} \xyoption{poly} \xyoption{knot}\xyoption{curve}
\input{diagxy}

\newcommand{\comment}[1]{}

\def\tn{\textnormal}

\def\mc{\mathcal}

\def\ZZ{{\mathbb Z}}

\def\RR{{\mathbb R}}

\def\Hom{\tn{Hom}}

\def\Ob{\tn{Ob}}

\def\to{\rightarrow}
\def\from{\leftarrow}
\def\cross{\times}
\def\taking{\colon}

\def\ss{\subset}

\def\iso{\cong}
\def\down{\downarrow}
\def\|{{\;|\;}}
\def\m1{{-1}}
\def\op{^\tn{op}}

\def\ullimit{\ar@{}[rd]|(.3)*+{\lrcorner}}
\def\urlimit{\ar@{}[ld]|(.3)*+{\llcorner}}
\def\lllimit{\ar@{}[ru]|(.3)*+{\urcorner}}
\def\lrlimit{\ar@{}[lu]|(.3)*+{\ulcorner}}
\def\ulhlimit{\ar@{}[rd]|(.3)*+{\diamond}}
\def\urhlimit{\ar@{}[ld]|(.3)*+{\diamond}}
\def\llhlimit{\ar@{}[ru]|(.3)*+{\diamond}}
\def\lrhlimit{\ar@{}[lu]|(.3)*+{\diamond}}
\newcommand{\clabel}[1]{\ar@{}[rd]|(.5)*+{#1}}

\newcommand{\arr}[1]{\ar@<.5ex>[#1]\ar@<-.5ex>[#1]}
\newcommand{\arrr}[1]{\ar@<.7ex>[#1]\ar@<0ex>[#1]\ar@<-.7ex>[#1]}
\newcommand{\arrrr}[1]{\ar@<.9ex>[#1]\ar@<.3ex>[#1]\ar@<-.3ex>[#1]\ar@<-.9ex>[#1]}
\newcommand{\arrrrr}[1]{\ar@<1ex>[#1]\ar@<.5ex>[#1]\ar[#1]\ar@<-.5ex>[#1]\ar@<-1ex>[#1]}

\newcommand{\To}[1]{\xrightarrow{#1}}

\newcommand{\Adjoint}[4]{\xymatrix@1{#2 \ar@<.5ex>[r]^-{#1} & #3 \ar@<.5ex>[l]^-{#4}}}

\def\Top{{\bf Top}}
\def\Cat{{\bf Cat}}

\def\Sets{{\bf Sets}}

\def\Pre{{\bf Pre}}

\def\colim{\mathop{\tn{colim}}}

\def\mcO{\mc{O}}

\def\mcS{\mc{S}}

\def\mcX{\mc{X}}

\newtheorem{theorem}{Theorem}[section]

\theoremstyle{remark}

\newtheorem{example}[theorem]{Example}

\newtheorem{question}[theorem]{Question}
\newtheorem{guess}[theorem]{Guess}

\theoremstyle{definition}

\def\Finm{{\bf Fin_{m}}}
\def\El{{\bf El}}
\def\Gr{{\bf Gr}}
\def\DT{{\bf DT}}
\def\DB{{\bf DB}}
\def\Tables{{\bf Tables}}

\begin{document}

\title{Table manipulation in simplicial databases}

\author{David I. Spivak}

\thanks{This project was supported in part by a grant from the Office of Naval Research: N000140910466.}

\begin{abstract}

In \cite{Spi}, we developed a category of databases in which the schema of a database is represented as a simplicial set.  Each simplex corresponds to a table in the database.  There, our main concern was to find a categorical formulation of databases; the simplicial nature of the schemas was to some degree unexpected and unexploited.

In the present note, we show how to use this geometric formulation effectively on a computer.  If we think of each simplex as a polygonal tile, we can imagine assembling custom databases by mixing and matching tiles.  Queries on this database can be performed by drawing paths through the resulting tile formations, selecting records at the start-point of this path and retrieving corresponding records at its end-point.  

\end{abstract}

\maketitle

\tableofcontents

\section{Introduction}

The distinguishing feature of the simplicial model for databases (see \cite{Spi}) is that the schemas are simplicial sets.  In other words, the organization of the data can be drawn as a picture consisting of vertices, edges, triangles, etc.  The purpose of this short note is to explain how the geometric aspect of such a schema can be directly useful for navigating data and manipulating tables.

There are two main applications we wish to emphasize at this time.  The first is the ability to add ``tiles" to an existing database to create a new one.  These new tiles (which are given as simplices) may come from internal or external sources.  For example, if a database has one section that involves people, and one section that involves US states, it might benefit from importing from an outside source a tile (1-simplex) that connects social security numbers to states of residence.  

The second is the ability to draw curves in a schema that indicate by what process one wishes to use data of one type to find corresponding data of another type.  As a heuristic example, imagine that one enters odometer readings at a location $A$, and draws a curve through a map beginning at $A$ and ending at $B$.  The system can be instructed to output a set of numbers corresponding to the expected readings of said odometers given travel along said route from $A$ to $B$.

In the following sections, we explain how to visualize simplicial databases, how to add tiles to create custom schemas, and how to use paths through a schema to indicate table manipulations.

While simplicial databases are purely mathematical objects, they can be visualized using a mathematical functor called ``geometric realization."  Throughout this paper, we imagine such a visualization, implemented on a modern computer.  Moreover, this implementation would allow the user to indicate aspects of the database using the computers mouse -- thus we may refer to clicking on simplices, dragging tiles, or drawing curves through a schema.

\section{Visualizing simplicial databases}

A simplicial database consists of a schema $X$ together with a sheaf of data $\mcO_X$; this is explained in \cite{Spi}.  In the present paper, we take the schema $X$ to be a symmetric semi-simplicial set, but we sometimes abuse terminology and call it a ``simplicial set" for short.  As such, it can be drawn on a user's screen by connecting together dots (0-simplices), edges (1-simplices), triangles (2-simplices), tetrahedra (3-simplices), and higher-dimensional tetrahedra ($n$-simplices, $n>3$).   

Simplices can only be connected along a common subsimplex.  For example, one cannot attach two triangles together along their spacious interior, or along part of some side -- only along vertices or edges. We associate to every simplex its set of attributes (vertex labels), and we can only glue two simplices together along a subsimplex if the labels match up in the obvious way.

The different simplices of a schema $X$ correspond to different tables in the database.  An $n$-simplex corresponds to a table with $n+1$ columns (one for each vertex), and with attributes specified by the labels of the vertices.  For example, a 2-simplex may correspond to a table with attributes ``First name," ``Last name," ``SSN."  The schema $X$ specifies how to connect these tables together.  For example, two triangles can be connected along a common edge or just a common vertex.
  
We do not know how best  to represent higher-dimensional simplices on a computer screen, so in the following discussion, we shall imagine that every simplex has dimension at most 3.  We might imagine a 3-simplex as rotating (confined only by its attachments to the rest of the schema).  Perhaps higher-dimensional simplices can be drawn simply as polygons or complete graphs.

The schema represents the table types, but does not ``have data in it."  Instead, the user should click on a simplex in the schema to see the corresponding table.  The separation between the schema and the data is represented mathematically as the separation between the simplicial set $X$ and the sheaf of data $\mcO_X$.  

Sometimes the data on a schema is only ``virtual" in the following sense.  Suppose we want to consider the operation of adding two integers together.  This can be represented as a virtual table with three columns, each with datatype ``integer."  In any row, the sum of the integers in the first two cells is the integer in the third cell.  Of course, we would never want to write down this entire table, but we ``know it exists" and can use it to compute.  In terms of the schema, it appears as a 2-simplex, but perhaps when one clicks it, the system displays the addition function itself and/or a few examples of it, rather than the entire virtual table.

\section{Adding tiles}

A company may have at its disposal many different sources of factual information, several of which come in the form of tables.  The company benefits when users have as much facility with these informative units as possible. 

The simplicial model offers the ability to quickly build custom databases by mixing and matching tiles.  A given user's need for information changes from moment to moment; he or she benefits from the ablilty to build up a database that will be most useful for the task at hand.  In the simplicial model, the  user can do so by selecting from the set of tables to which he or she has access, each of which is represented as a triangle, edge, etc.  The user drags various tiles down from the library to his or her own workspace and connects them together so as to enable quick and flexible queries.  The idea of a single database for the whole company may start to seem rigid and old-fashioned. 

Tiles may be color-coded, hidden from view, available in suites, etc.  There is plenty of room for innovation here.  For example, the color on older tables may fade as the data loses its freshness, verified tables may be indicated with a check-mark ($\checkmark$), questionable information may be flagged, etc.

Sometimes, one may wish to get data from an outside source.   Such tiles could be made available for purchase.  It should not be hard to keep track of where each tile came from and when it was created.  In this sense, we offer a solution to the ``data provenance" problem.  We imagine that companies would visually trademark their tiles in an effort to ensure quality and prevent fraud.  

On the opposite end of the spectrum, some tiles may simply be well-known mathematical operations such as addition.  If a table in $\mcX$ has two columns we wish to add together, we can simply attach an ``addition tile" to $\mcX$ by connecting its ``summands" edge to the edge in $X$ representing the two columns in question.  In Section \ref{sec:curves} we shall see how the summation can be performed by drawing a curve through the ``summands edge" and ending at the ``sum" vertex of the addition tile.  

\begin{example}

Here we present an example that theoretically fits into the same mold, but is somehow different in that we are gluing a vertex along a vertex.  

Suppose we have a tile $A$ in which one of the vertices is of type ``date."  We always have available the one-row, one-column table ``today's date."  As a schema, it looks like a single vertex.  We can drag the ``today's date"  vertex to the date vertex of $A$ and drop it; the result will be a tile with date replaced by ``today's date."  The data over that simplex will be the result of selecting only those rows whose date is that of today.

\end{example}

\begin{example}

Each individual may have his or her own identity in the form of a tile.  This tile may have attributes such as ``First name," ``Last Name," etc., but as a table it (probably) only has one row.  Again, this tile may be dragged and dropped into an existing schema.

For example, suppose one has a ``birth" tile that includes all the data of the person's birth (date, time, location, etc.).  This tile could be dropped into a ``zodiac" schema with these vertices as its ``input" (boundary).  Doing so has the effect of selecting all the pertinent data about planetary and stellar locations at that time.  Using methods developed in the next section, we can create a tile with precisely this data.  Next, one could plug that tile into their choice of ``horoscope" schemas to return an associated text. 

\end{example}

\begin{example}

Suppose one drags a tile of a given type onto a tile of the same shape.  If we apply the above ideas to this situation, the result will basically be the intersection of the two tables (if the tables have repeated rows, the construction will in fact result in the {\em fiber product} of the two tables).  

One might instead wish to take the union of the two tables in question.  This could easily be done, For example, whenever one drags a tile and places it as a subtile of an existing one, perhaps the machine could ask whether the user is requesting a union of intersection.  If it is a union, the machine would ask for additional information: do a UNION ALL, a forgetful UNION, or something more controlled (as in \cite{Spi}).

\end{example}

\section{Drawing paths}\label{sec:curves}

In the introduction, we mentioned that if one has a table of odometer readings and a route through a map, he or she should be able to output odometer readings expected after taking the indicated journey.  This is literally an application of the following more general procedure.  We give it as Example \ref{ex:odometer}.

Suppose that the user has two tables: $T_1$ has attributes $A,B$, and $T_2$ has attributes $B,C,D$.  The user wants to query the database in the following way.  Given a list of data of type $A$, he wants to select all data in $T_1$ that conform, then use them to query table $T_2$ and locate all the corresponding data of type $C,D$.  For example, given a last name $\ell$, find all the social security numbers that correspond to $\ell$, and return the set of incomes and withheld incomes associated to each.

This query might take a SQL expert a few minutes to construct, but with simplicial databases it's quite easy.  Recall that the schema for the above situation consists of an edge and a triangle; the edge has vertices labeled $A$ and $B$, the triangle has vertices labeled $B,C$, and $D$, and the two are attached at $B$. Once the user has chosen a set of data of type $A$, he simply clicks the schema at the vertex $A$, and drags a curved line through the schema that ends at the 1-simplex $C,D$.  A good implementation of simplicial databases can then return the desired data of type $C,D$. 

\begin{example}\label{ex:odometer}

One can consider a map $M$ as the schema for a simplicial database, all of whose simplices have dimension 0 or 1.  The vertices of $M$ are the intersections of roads and the 1-simplices are the road-portions between these intersections.  Each road-portion has a distance in miles, which we can record as an integer $d$ for this discussion.  If each road portion is a 1-simplex in the schema, then over each road portion we need a table with two columns.  We use the table of all pairs of integers $s,t$ where $t-s=d$. 

Now, suppose we have a 1-column table $T$ of odometer-readings.  It will be a subtable of the 1-column table over any intersection $A$ of the map.   One can then draw a route through the map from $A$ to $B$.  Using the above techniques, this will result in a functor from tables over $A$ to tables over $B$.  Applying this functor to $T$ will result in a table over $B$ whose entries are the integers corresponding to the new odometer readings given this journey.  

\end{example}

\begin{example}

Suppose we keep track of all interactions between pairs of companies and the date each occurred on.  The data consisting of which two companies interacted and on what date is a 2-simplex (3 columns).  Suppose we also keep track of when two companies join forces to create something together.  This is also represented with a 2-simplex: which two companies are interacting and what they create together.  

Given access to these two tiles, a user might decide to attach them along the common face.  The resulting shape is a rhombus.  What is the meaning of drawing a curve through this rhombus, beginning at ``date" and ending at ``common creation"?  This curve represents a functor from category of sets of dates to the category of sets of common creations.  

This functor is computed as follows.  Suppose given a date (resp. set of dates).  Begin by selecting all interactions between companies that occurred on that date (resp. set of dates).  Some of these interactions will have resulted in ``common creations."  Finish by outputting this set of common creations.

In other words, this path through the schema represents the query ``tell me all the common creations that occurred on these dates."  

\end{example}

\begin{example}

Suppose we have a 1-simplex tile in which we have a list of friendship pairs.  In other words, over each vertex is a set of people, and over the 1-simplex are those pairs of people that are friends.  

We can grab one vertex of the 1-simplex and drop it onto the other vertex, creating a loop.  This will have the effect of intersecting the set of people represented by the first vertex with the set of people represented by the second vertex.  It will also ``throw away" friendships if their participants are not in both sets of people.  (In general, we are actually taking a fiber product, not an intersection.  However, if there are no duplicates, this fiber product is nothing more than an intersection.)

We can draw a curve through this new loop, say beginning at the vertex, going around the loop three times, and ending at the vertex.  Given a list of people $L$ at the first vertex, the above curve results in the query ``name everyone who is a friend of a friend of a friend of a person in list $L$."    

\end{example}

Different paths through a schema may result in different queries, and this may be a good thing.  However, sometimes one wants to know which pairs of paths are guaranteed to return equal queries.  It should not be a difficult problem to determine when that is the case and what kinds of constraints can control for it.  This is a kind of homotopy problem, and may have topological solutions.  These solutions will be dependent on the sheaf of data $\mcO_X$, not just on the shape of the schema $X$.

\appendix

\section{Technical details}\label{sec:technical}

In this section we give the technical details of the above constructions.  None of them are hard; the innovation here is in noticing their existence not in the mathematics underlying them.  In the first subsection below, we give a very brief overview of simplicial sets (or more precisely, symmetric semi-simplicial sets).  In the next two subsections we explicate the category theory behind the concept of adding tiles and that of drawing curves to represent queries.

\subsection{Visualizing simplicial databases}

Let $\Finm$ denote the category of finite non-empty sets and the monomorphisms between them.  The category of semi-simplicial sets is $\mcS:=\Pre(\Finm)$, so that an object in $\mcS$ is a functor $X\taking\Finm\op\to\Sets$ and a morphism of semi-simplicial sets is a natural transformation.

As with any presheaf category, every object $X\in\Ob(\mcS)$ has a {\em category of simplices}, which we denote $\El(X)$.  Its objects are the simplices of $X$ and its morphisms are the inclusions of simplices.  It can be described precisely as the opposite of the Grothendieck construction, $\El(X)=\Gr(X)\op$.  Recall that the Grothendieck construction of $X$ is the category whose objects are pairs $(A,x)$ where $A\in\Finm$ is a finite non-empty set and $x\in X(A)$ is an element in the set $X(A)$.  A morphism $(A,x)\to(A',x')$ in $\Gr(X)$ is a monomorphism $f\taking A'\to A$ such that $f(x)=x'$. 

One can visualize an object $X\in\mcS$ as the union of its simplices.  In other words, there is a functor $\El(X)\to\Pre(\Finm)$ called the {\em diagram of simplices} of $X$ and we shall soon see that $X$ is the colimit of its diagram of simplices.  We will now explain this idea.  There is a Yoneda functor $y\taking\Finm\to\mcS$, sending $A$ to $yA:=\Hom_\Finm(-,A)$, which is fully faithful.  Given an object $(X\taking\Finm\op\to\Sets)\in\mcS$ and a finite set $A$, one has $X(A)=\Hom_\mcS(yA,X)$.

If one considers every object in $\mcS$ to be a ``formal union of objects in $\Finm$", then the Yoneda imbedding realizes each object of $\Finm$ as the union of just itself.   In general, any object $X\in\mcS$ is the colimit of its diagram of simplices: \begin{align}\label{dia:iso}X\iso\colim_{yA\to X} yA.\end{align}  

One can visualize the Yoneda image of any finite non-empty set $A$ as the ``polygonal hull" of $A$ as a set of vertices: a one-point set is seen as a vertex, a two-point set is seen as a line-segment, a three-point set is seen as a triangle, a 4-point set is seen a tetrahedron, etc.  The isomorphism \ref{dia:iso} displays $X$ as the union of these basic shapes.

This visualization can be made precise in the following sense.  There is a ``geometric realization" functor $\Re\taking\mcS\to\Top$, where $\Top$ is the category of topological spaces.  It is constructed as follows.  Any finite non-empty set is isomorphic to $\{1,2,\ldots,n\}$ for some $n\in\ZZ_{\geq 1}$.  The $n$-simplex $\Delta^n$ can be realized in $\Top$ as the subspace of $\RR^{n+1}$ whose points are $$\Delta^n:=\Re(yA):=\{(x_0,x_1,\ldots,x_n)\in\RR^{n+1}_{\geq 0}|x_0+x_1+\cdots+x_n=1\}$$  Morphisms of finite sets can be linearly extended to continuous functions between these spaces.  To complete the construction, given any $X\in\mcS$, we set $$\Re(X)=\colim_{yA\to X}\Re(yA),$$ a formal union of topological simplices.

We can alter the above without making a significant difference.  Instead of looking at $\Finm$, we consider the comma category $(\Finm\down \DT)$ for a given set $\DT$ of data types.  We set $\mcS_\DT=\Pre(\Finm\down\DT)$.  The objects of $\mcS_\DT$ are simplices as above, except that every vertex is labeled by a data type.  A morphism of $\mcS_\DT$ is also as above, with the added requirement that vertices are sent to vertices of the same $\DT$-label.

A simplicial database $\mcX$ is more than just a simplicial set $X$ with labeled vertices -- it also has a sheaf of data $\mcO_X$.  Each simplex $x\in X(\sigma)$ represents a table $\mcO_X(x)$ whose attributes are $\sigma(A)\ss\DT$.  All this is made more precise in \cite{Spi}.

\subsection{Adding tiles}

Certain joins of databases can be done by simply ``dragging and dropping tiles."  In other words, given two tiles $(\sigma_1,\tau_1)$ and $(\sigma_2,\tau_2)$ that share a common face $(\sigma,\tau)$, we can connect them together without additional thought as follows.

For any simplex $\sigma\taking A\to\DT$, there is a universal table on it, which in the parlance of \cite{Spi} is $\Gamma(\sigma)$.  The limit of the diagram $$(\sigma_1,\tau_1)\to(\sigma,\Gamma(\sigma))\from(\sigma_2,\tau_2)$$ can be taken in the category of simplicial databases $\DB$, and the result is two simplices connected along $A$.

The above construction can always be done.  Sometimes, instead of using the universal table on $A$, one may opt for a more controlled table on $A$.  This is detailed in \cite{Spi}.  In terms of visualization, one may imagine that the tiles have additional markings on their $A$-faces to indicate compatibility.  These markings in fact indicate the specific table along which we are joining.

\subsection{Drawing paths}

Drawing a continuous path through a picture of a (symmetric semi-) simplicial set $X$ can be considered a zig-zag of morphisms in the category $\Gr(X)$.  The starting point of the curve is an object in $\Gr(X)$.  As one draws the curve, it passes from a simplex to one of its faces or vice-versa.  Each time it moves from a simplex to one of its faces, there is a corresponding morphism in $\Gr(X)$.  Each time it moves from a face to a bigger simplex, there is a morphism ``going the opposite way" in $\Gr(X)$.  Thus this path results in a zigzag of arrows in $\Gr(X)$.

For example, consider two triangles $ABC$, $CDE$ attached along a vertex.  We begin at the side $AB$, run through the interior of the first triangle, exit through vertex $C$ and travel along $CD$ until we get to $D$.  This gives the zigzag $$``AB"\from ``ABC"\to ``C"\from ``CD"\to ``D."$$ 

Thus, to explain the construction we discussed in Section \ref{sec:curves}, we need to give functors $F\taking\Gr(X)\to\Cat$ and $G\taking\Gr(X)\op\to\Cat$ that agree on objects.  These are developed in \cite{Spi}, but we need to explain here.

Given a simplex $\sigma\in\Gr(X)$, assign $F(\sigma)=G(\sigma)=\Tables_\sigma$.  If $i\taking\sigma\ss\sigma'$ is a subsimplex, we need functors $F(i)\taking\Tables_{\sigma'}\to\Tables_\sigma$ and $G(i)\taking\Tables_{\sigma}\to\Tables_{\sigma'}$.  These are the pullback and push-forward functors $F(i):=i^*$ and $G(i):=i_+$. 

To be explicit, suppose we have $i\taking\sigma\ss\sigma'$ as above and a table $K'\To{\tau'}\Gamma(\sigma')$.  The functor $F(i)$ sends $\tau'$ to the bottom row in the commutative diagram $$\xymatrix{K'\ar[r]^{\tau'}\ar@{=}[d]&\Gamma(\sigma')\ar[d]^{i^*}\\K'\ar[r]_{i^*\circ\tau}&\Gamma(\sigma).}$$  The functor $F(i)$ acts similarly on morphisms of tables: simply replace both instances of $K'$ above with a morphism $K'_1\to K'_2$.

The functor $G(i)$ sends a table $K\To{\tau}\Gamma(\sigma)$ to the bottom row in the pullback diagram $$\xymatrix{K\ar[r]^\tau&\Gamma(\sigma)\\K\cross_{\Gamma(\sigma)}\Gamma(\sigma')\ar[r]_-{i^*\tau}\ar[u]\lllimit&\Gamma(\sigma')\ar[u]_{i^*}}$$  The functor $G(i)$ acts similarly on morphisms of tables: $$\xymatrix{K_1\ar[r]&K_2\ar[r]^\tau&\Gamma(\sigma)\\K_1\cross_{\Gamma(\sigma)}\Gamma(\sigma')\ar[r]\ar[u]\lllimit&K_2\cross_{\Gamma(\sigma)}\Gamma(\sigma')\ar[r]_-{i^*\tau}\ar[u]\lllimit&\Gamma(\sigma')\ar[u]_{i^*}}$$

Given $F$ and $G$, we can now construct a functor from the category of zigzags in $\Gr(X)$ to the category of categories, as above.  It is this construction that we use to query the database.  Given a path through the schema, we get a zigzag in $\Gr(X)$ as above, then apply $F$ and $G$ to get a functor from tables over the start-point to tables over the end-point.  

We should record one more important point.  As we have been saying, a path through $X$ induces a zigzag of tables $$(\sigma_1,\tau_1)\to(\sigma_2,\tau_2)\from(\sigma_3,\tau_3)\to(\sigma_4,\tau_4)\from(\sigma_5,\tau_5)$$ in the form of projects and selects.  Write $\tau_1\taking K_1\to\Gamma(\sigma_1)$, and suppose that $K_1'\to K_1$ is a table over $\tau_1$.  Applying the above constructions we get a diagram $$\xymatrix{K'_1\ar[d]&\stackrel{K'_2:=}{K'_1}\ar[d]\ar@{=}[l]&\stackrel{K'_3:=}{(K'_2\cross_{K_2}K_3)}\urlimit\ar[l]\ar[d]&\stackrel{K'_4:=}{K'_3}\ar[d]\ar@{=}[l]&\stackrel{K'_5:=}{(K'_4\cross_{K_4}K_5)}\urlimit\ar[d]\ar[l]\\K_1\ar[r]\ar[d]^{\tau_1}&K_2\ar[d]^{\tau_2}&K_3\ar[d]^{\tau_3}\ar[l]\ar[r]&K_4\ar[d]^{\tau_4}&K_5\ar[d]^{\tau_5}\ar[l]\\\Gamma(\sigma_1)\ar[r]&\Gamma(\sigma_2)&\Gamma(\sigma_3)\ar[r]\ar[l]&\Gamma(\sigma_4)&\Gamma(\sigma_5)\ar[l]}$$ in which $K_2,\ldots,K_5$ are induced either as equalling the previous, or by fiber product.  What we have not yet said is that from this, we can construct a function $K_5'\to K_1'$.  Such a morphism will exist for any zigzag in $\Gr(X)$.

Finally, we have maps $K_5'\to\Gamma(\sigma_5)$ and $K_5'\to K_1'\to\Gamma(\sigma_1)$, which induce a single map $K_5'\to\Gamma(\sigma_1)\cross\Gamma(\sigma_5)$.  This map gives a table with rows $\sigma_1\amalg\sigma_5$ and is the graph of the functor given by the zigzag applied to our starting table $\tau_1$. 

\bibliographystyle{amsalpha}

\end{document}